# Quantum State Absorptions Coupled To Resonance Raman Spectroscopy Could Result In A General Explanation of TERS

Zachary D. Schultz[1]*, John Parthenios[2]*, Rimma Dekhter[3], Dimitris Anestopoulos[2], Spyridon Grammatikopoulos[2], Kostantinos Papagelis[2,4], James M. Marr[1], David Lewis[3], Costas Galiotis[2,5], Dimtry Lev[6] and Aaron Lewis[6]*^

1. Department of Chemistry and Biochemistry, University of Notre Dame, Notre Dame, Indiana 46556, USA
2. FORTH/ICE-HT, Stadiou Street, Platani, GR-265 04, Patras, GreeceFoundation of Research and Technology Hellas, Institute of Chemical Engineering Sciences, 26504 Patras, Greece
3. Nanonics Imaging Ltd, Jerusalem, 91487, Israel
4. Department of Materials Science, University of Patras, 26504 Patras, Greece
5. Department of Chemical Engineering, University of Patras, 26504 Patras, Greece
6. Department of Applied Physics, The Hebrew University of Jerusalem, Jerusalem 9190401, Jerusalem

Abstract

Tip enhanced Raman scattering (TERS) amplifies the intensity of vibrational Raman scattering by employing the tip of a probe interacting, in ultra close proximity, with a surface. Although a general understanding of the TERS process is still to be fully elucidated, scanning tunneling microscopy (STM) feedback is often applied with success in TERS to keep a noble metal probe in intimate proximity with a noble metal substrate. Since such STM TERS is a common modality, the possible implications of plasmonic fields that may be induced by the tunneling process are investigated and reported. In addition, TERS of a 2D resonant molecular system, a $MoS_2$ bilayer crystal and a 2D non-resonant, lipid molecular bilayer is compared. Data with multiple excitation wavelengths and surfaces for the resonant system in the near- (TERS) and far-field regimes are reported. An interpretation based on weak coupling interactions within the framework of conventional resonance Raman scattering can explain the observed TERS enhancements. The non-resonant molecular lipid system, on the other hand, requires strong coupling for a full understanding of the reported observations.

**Introduction**

Raman scattering is a widely applicable material characterization technique. Nonetheless, its resolution has been limited. As optical methods proliferate with nanometric resolution, there is a great desire to extend Raman imaging into the nanometric regime. Tip enhanced Raman scattering (TERS) [1-6], where a tip of a scanned probe microscope enhances the Raman spectrum of a molecule on a surface, has generated hope that this can be achieved. Recently some spectacular examples of the application of TERS at the single molecule and sub-single molecule level have been published [7-12]. Although these latter publications have all been achieved by using scanning tunneling microscopy as the feedback mechanism of a gold tip on a gold surface, there are many emulations of TERS with a variety of feedback mechanisms and a variety of combinations of tip and sample substrate.

A technique related to TERS is the method known as surface enhanced Raman scattering (SERS) [13] that also has been extensively investigated over the last 40 years. However, both of these effects have eluded a general explanation and debates remain about the origin of the enhanced signals seen in TERS and SERS.

Of these two related effects TERS in principle could be considered to have more experimental flexibility than SERS and has matured sufficiently to provide a platform that could enrich our understanding of the origins of these Raman enhancement phenomena. Our investigations have employed several different directions of experimental measurement to broaden the understanding of the TERS and hopefully SERS effect.

From a general perspective, noble metals, commonly associated with plasmon generation, are often used for TERS tips and substrates. In addition to this scanning tunneling microscopy (STM) feedback between a noble metal tip and surface is a prevalent modality in TERS for bringing and keeping the probe in ultra close proximity to the surface on which the molecular system, whose Raman spectrum is to be enhanced, has been placed.

Several questions arise from the success of STM in TERS. First, tunneling by itself can generate all k vectors and all energies and thus could lead to the very effective generation of plasmons in the noble metal tip/substrate gap even without excitation of the Raman laser. This possibility is one of the questions addressed experimentally and reported in this paper. In addition, the effects of imposed voltage and current that are a part of STM feedback are also considered in terms of the observed TERS results.

These investigations and their conclusions are then analyzed by studying two extreme 2D systems with TERS. One is a molecular system resonant with some excitation wavelengths used and the other is a system non-resonant with the excitation wavelength employed.

For the resonant molecular system, a $MoS_2$ bilayer crystal is explored by studying the wavelength dependence of the resonance Raman enhancement in the far-field in and out of resonance as contrasted with what is observed in the near-field. This is compared for a variety of tip, sample and substrate choices. Such alterations have never been consistently investigated with TERS. The question to be addressed in this sequence of studies is what are the alterations induced by the presence of the tip on the well-documented resonance Raman behavior of $MoS_2$.

With these resonant results, a non-resonant lipid bilayer 2D system, dipalmitoyl phoshatidylcholine (DPPC) deposited on mica is investigated with TERS enhancement. The TERS observations on such a non-resonant system are distinct from what is observed on resonant systems.

The ultimate goal of our paper is to integrate the observations in this paper with reports in the literature in order to evolve a general hypothesis that can explain the phenomenon of TERS.

**Background**

Typically TERS spectra are recorded in a somewhat arbitrary fashion without the luxury of predictive analysis. A particular probe/material/substrate combination with a certain excitation wavelength leading to effective enhancement of the signal has been imposed by what produces a result. Rarely has the question been asked as to how the results may change with different conditions of excitation wavelength, sample surfaces, etc. for the same probe.

Nonetheless, it should be noted that three recent papers stand out in this regard. In one paper the effect on observed TERS results of tuning the plasmonic absorption in the tip by Focused Ion Beam (FIB) milling was examined at one wavelength [14]. This approach is tip centric and could be categorized within the extensive literature on what should be the plasmonic resonance of the tip material based on the noble metal used and the geometry of the tip.

Similar to the above study the group of Kawata and Verma has also investigated TERS probes with multiple nanoparticles at the tip. As a function of the number of particles, the plasmon resonance changes. These authors chose a Raman excitation wavelength for TERS within this resonance [15]. It should be noted that the molecular system investigated (carbon nanotubes) also had a resonance close to the plasmon resonance of the tip. Nonetheless, the tip did indeed affect the enhancement of the molecule. However, in this case the experiment was not repeated with other wavelengths.

Within this group of papers is the excellent study by Chiang et al [12] who investigated with STM in ultrahigh vacuum the spectra of a porphoryn monolayer at molecular resolution. Both TERS and tip enhanced fluorescence were investigated and four distinct laser lines were chosen to study this monolayer with a silver tip and a silver surface. The four laser lines corresponded to the resonant conditions of the 4 porphoryn Q bands for each of the two lowest electronic excited states of the porphyrin. No out of resonance wavelengths were investigated.

In studies such as these, as exemplified by Chiang et al [12] or Zhang et al [7] where single molecules are studied, it is common to use an approach where a noble metal tip with plasmonic properties and a noble metal surface are chosen to form what has now been called a gap mode. This has overtones of emulating the observations in SERS where field enhancing nanojunctions between plasmonic nanoparticles are thought to occur [16]. Indeed experiments with AFM controlled noble metal tips above nanoparticles of gold show the expected enhancements [17-18].

In addition to these studies there is an increasing body of literature that reports not only TERS on non-noble metal surfaces as has been reported for some time, [19] but, also SERS surfaces that do not have noble metal character [20,21].

**Methods**

*Tunneling Based Plasmon Production*

A repeated TERS geometry employs STM tunneling feedback between a gold tip and a gold surface.  One experimental direction in this paper is aimed at delineating if such a tunneling gap could be independently producing plasmons and their associated fields. To investigate such questions an AFM probe with tuning fork based feedback, which allows for switching from AFM to STM feedback, is combined with a near-field scanning optical microscope (NSOM) system in a multiprobe scanned probe geometry. As shown in Figure 1  an AFM/STM gold probe establishes contact with a gold surface and is maintained in feedback with either AFM and STM. Simultaneously, a second NSOM probe monitors the near-field optical signal either using apertured or scattering NSOM (Nanonics MultiProbe MultiView 4000, Nanonics Imaging Ltd, Jerusalem, Israel). Relevant to this is that the form of probes used in our study have been shown, in a multiprobe system, to be able to be brought into physical contact with one another [22].  For the experiments reported in this paper an apertured NSOM probe was used with the AFM/STM probe in order to eliminate any chance of photons not propagating along the gold surface from interfering with the detected signal.

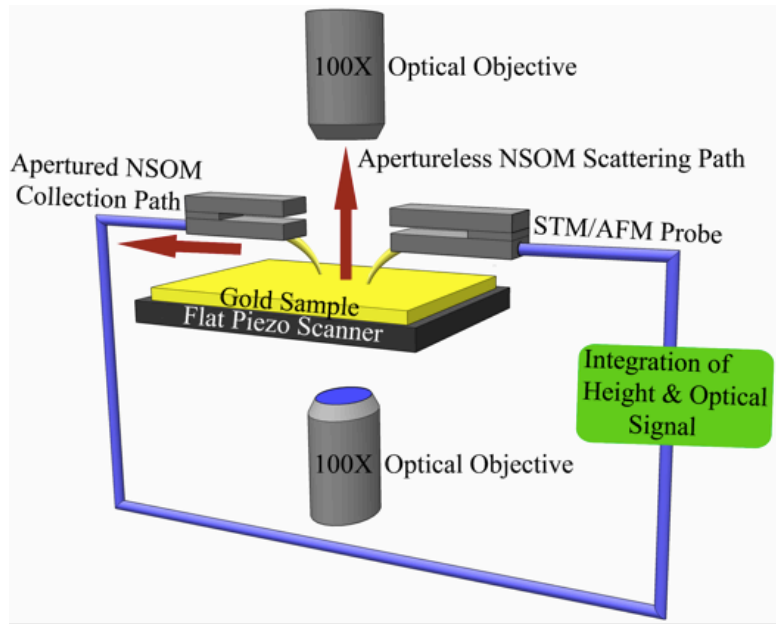

Figure 1.  An illustration showing the multiprobe scanned probe configuration for understanding tunneling induced plasmon production on a gold surface.  A diagrammatic illustration of an AFM/STM probe in tunneling feedback with a gold surface is shown with a second NSOM probe that is used to collect, as a function of distance from the tunneling probe, evanescent plasmons generated by the STM.

*TERS Measurements*

The probe tip used in the TERS experiments reported are composed of a heterogeneous material combining a noble metal in a dielectric environment.  The probes (Nanonics Imaging

Ltd, Jerusalem, Israel) have a completely transparent probe shaft to minimize scattering effects [23].

The geometry used was top down illumination of the probe tip appropriately cantilevered. The bending angle was chosen so that no external ray of the objective lens from above was obscured. This is illustrated in Figure 2. These probes and such an illumination geometry allowed for effective data collection on opaque samples when the probe tip was placed in the Z polarization lobe of a Gaussian focused laser beam.

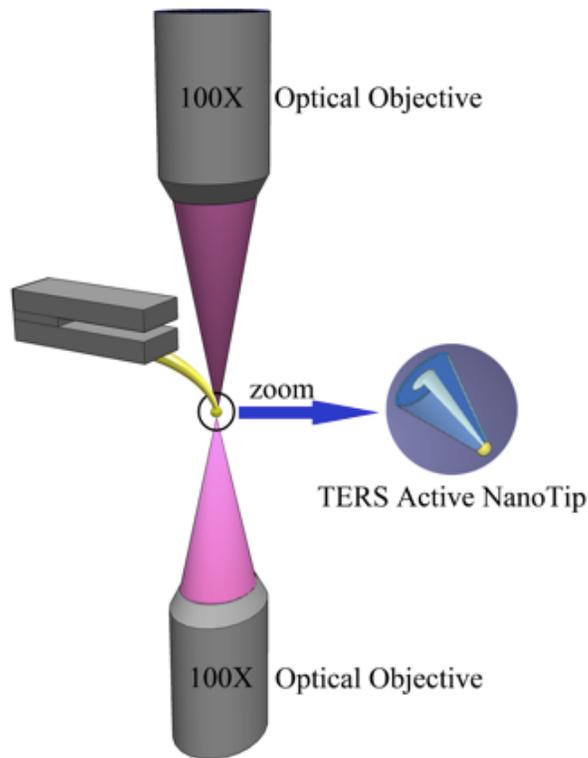

**Figure 2.** TERS active cantilevered glass probes with extended tip and completely transparent probe shafts enable a top illumination geometry with maximum tip enhancement and minimum background scattering from the tip shaft and cantilever.

*AFM Feedback Methodology*

As part of these investigations TERS were applied with normal force tuning fork feedback. Such feedback does not impart any background illumination from a feedback laser. In addition, tuning fork probes do not exhibit jump to contact which is ubiquitous in beam bounce feedback and thus can be kept in close proximity to the sample even permitting switching between AFM and tunneling feedback.

*TERS Investigation of $MoS_2$*

TERS was performed on mechanically exfoliated $MoS_2$ bilayer crystals supported on a polymeric substrate or silicon. The bilayer $MoS_2$ crystals were exfoliated from bulk $MoS_2$ (MN Nanomaterials, UK) onto either a thin SU8 polymer film (Shipley Photoresists Inc.) which had been spin coated on a SiO$_2$ substrate or onto silicon. In addition, CVD grown MoS$_2$ on a glass substrate was also investigated.

TERS data were recorded using a Nanonics MultiView 2000 (Nanonics Imaging Ltd., Jerusalem, Israel) with a TERS probe as described above. The AFM system was integrated with an inVia reflex Raman spectrograph (Renishaw, Wotton-under-Edge, England) using an Ultra long working distance Olympus 50X, 0.45 NA objective. Excitation was achieved with a 514.5nm argon ion laser, a 532nm doubled Nd;YAG laser and a 785 nm diode laser. Acquisition times for all spectra were approximately 1sec.

*TERS Investigations of Dipalmitoyl Phoshatidylcholine Lipids*

Tip enhanced Raman scattering was performed using a Nanonics MV4000 with a TERS probe as described above with the AFM incorporated into a custom Raman microscope. A 633 nm (HeNe) laser illuminated the TERS probe from above. The TERS spectra were obtained by averaging 10, 1sec acquisitions. The objective used was an Olympus 40x water immersion objective (LUMPlanFLN, 0.80 N.A.). Supported lipid bilayers were formed on freshly cleaved mica using previously reported methods [24].

Briefly, single uni-lamellar vesicles were formed by hydrating 10 mg of dipalmitoyl phoshatidylcholine lipids in 1 mL of water (18.2 MΩ cm resistivity). [25]. The lipids were subjected to five freeze-thaw cycles prior to extrusion through a 1μm pore track-etch polycarbonate membrane. The resulting vesicles were then serially extruded through smaller pore membranes to a final size of 100 nm. Vesicles solutions were added dropwise onto freshly cleaved mica and supported bilayer membranes formed from spontaneous vesicle fusion and rupture on the surface.

**Results**

*Plasmon Production in the STM Interactions of a Noble Metal Probe With a Noble Metal Surface*

The possibility of plasmonic excitation and propagation on a noble metal surface with scanning tunneling feedback has been known for many years [26,27]. These studies generally have used a tunneling tip to induce tunneling on a gold surface and prism out-coupling of plasmonic emission from the bottom of the thin film surface.

In contrast the experiments reported in this paper with a multiprobe scanned probe microscope allow the detection of plasmons generated by an STM tip on the same top side of the surface where a molecule would sit in a TERS geometry (see Figure 1). This is accomplished with an apertured NSOM probe acting as the collecting second probe. In fact, as can be seen in Figure 3, that it has been possible to demonstrate plasmon propagation in close proximity to the tunneling tip.

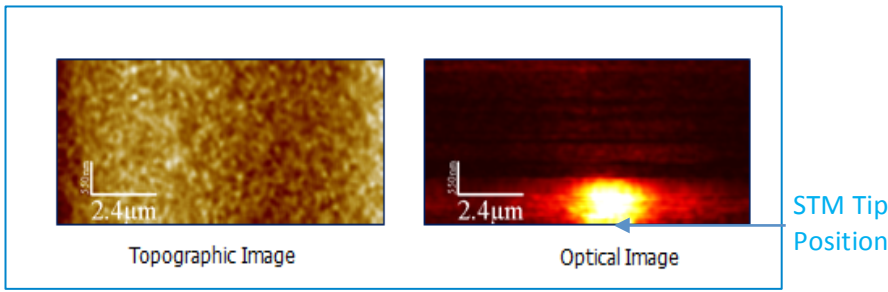

Topographic Image     Optical Image     STM Tip Position

Figure 3: Simultaneously obtained topography and near-field optical collection mode images of plasmon propagation on a gold surface from the point of their excitation by an STM tip in the lower middle of the right image indicated by a blue arrow. The STM probe tip was held at a constant point to excite plasmons and the NSOM probe was scanned from this point to form the images shown as a function of distance from the STM probe.

The conditions of plasmon production as monitored by the NSOM probe are seen in Figure 4. To detect plasmons, the voltage of the tunneling junction had to be increased above 1.5 V. As shown in Figure 4, a higher voltage generates a higher optical signal. It should be mentioned in this regard that, when the feedback was switched to AFM feedback and the same voltage was imposed between the probe and the sample there was no optical effect. Thus, to see plasmon production the tunneling current was required and this is understandable since such a tunneling current produces all k vectors with all energies for very effective excitation of plasmons.

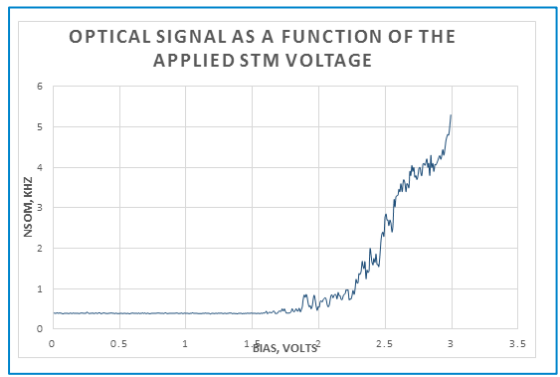

Figure 4. Collected optical signal as a function of bias voltage between the gold STM tip and the gold substrate. The tunneling current was on the order of nanoAmps.

However, comparing these results with the TERS data of Zhang et al [7] demonstrates that neither the voltage used of 125mV or the tunneling current of tens to hundreds of picoAmps (pA) at constant voltage are consistent with our results which required 1.5 volts and a current of nanoamps (nA) for plasmon generation. Thus, the STM process by itself certainly does not aid plasmon generation in the STM/molecule/surface gap. In spite of this, the results of Zhang et al[7] do clearly show that tunneling current is related to the intensity of the TERS signal. Specifically, the higher the tunneling current the larger the intensity of the TERS signal. This is shown in Figure 5 which shows that as the current increases the tip distance from the surface decreases and the TERS signal increases.

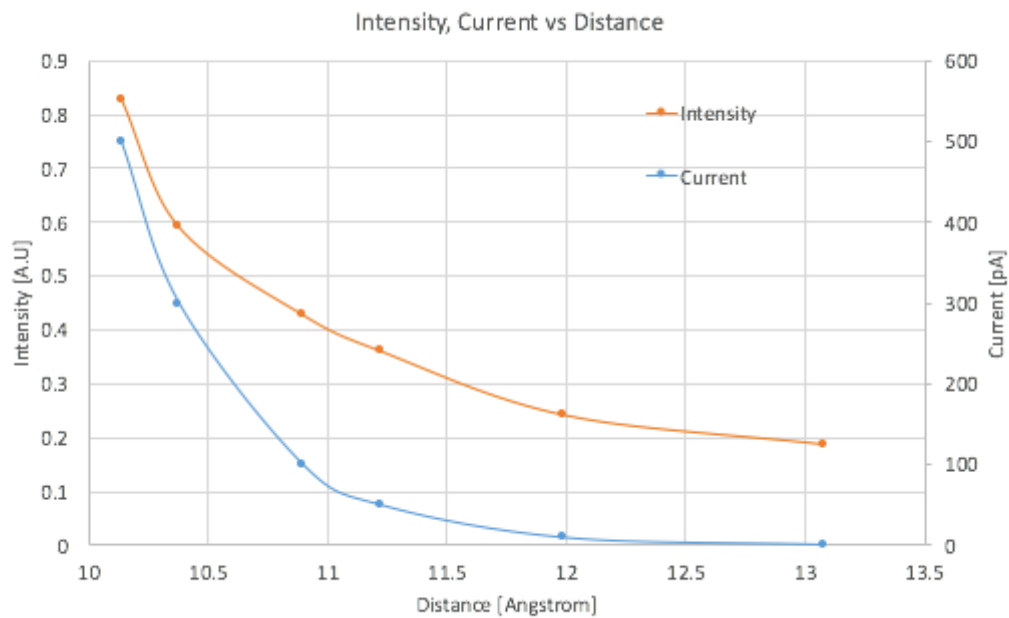

Figure 5. Data from Zhang et al [7] have been graphed as a function of Raman intensity versus distance. The construction of the graph from the data of Zhang et al [7] is described in the supplementary material associated with this paper.

*TERS Investigations of MoS$_2$ As A Function of Laser Excitation Wavelength and Substrate*

TERS and far-field Raman results of a bilayer crystal of MoS$_2$ exfoliated on polymer and silicon substrates with different excitation wavelengths are shown in Figure 6 A and B. These results are described within the context of a recent report of a TERS study with 660nm excitation of a 4 layer MoS$_2$ sample sandwiched between a gold tip and gold surface [28]. The 660nm excitation used in this previous study [28] is resonant with the A direct exciton band as deduced from the absorption of MoS$_2$ reported by Dhakal et al [29]. In contrast the 532nm and 514nm excitation data reported in Figure 6 are within a rising absorption background that peaks at around 442nm. The 785nm excitation in Figure 6A is not resonant with any absorption band in the spectra reported by Dakal et al [29] which are essentially the same for single to a few layers of MoS$_2$. Thus, the data in Figure 6A and B over many wavelengths, some resonant and others not, allow for a delineation of the contrasting effects of near-field TERS with far-field resonance Raman enhancements.

Figure 6A

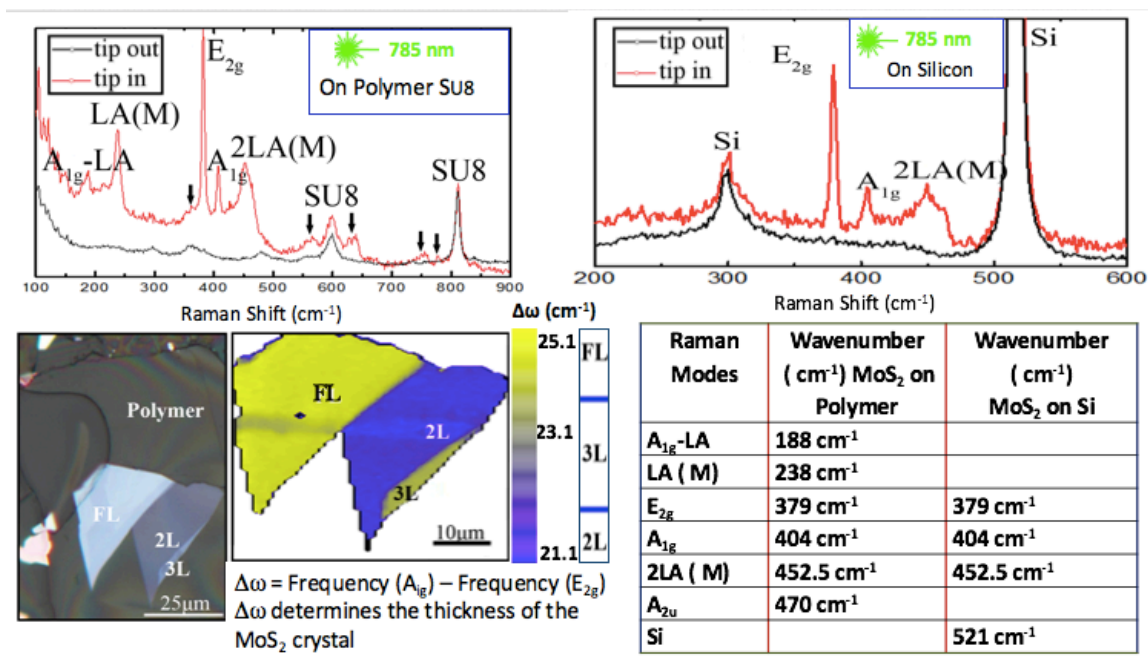

Figure 6B

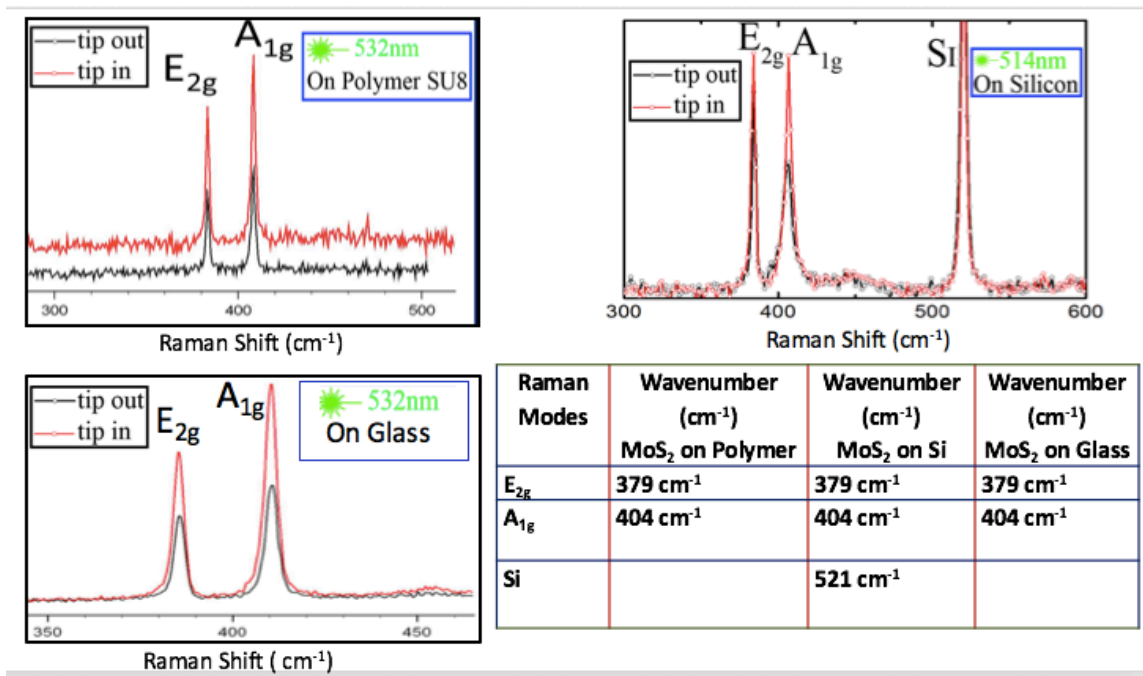

**Figure** 6A. Far-field and TERS on a bi-layer MoS$_2$ film supported on a SU8 polymer and silicon substrate excited with 785 nm excitation. Figure 6B. Far-field and TERS on a blilayer MoS$_2$ on SU8 and silicon excited with 532nm and 514 nm lasers respectively. The far-field and TERS

results on CVD grown $MoS_2$ on a glass substrate with 532nm excitation is also shown for comparison. Laser powers of 6mW (785nm), 0.9mW (532nm) and 0.3mW (514nm) were employed respectively. A 50X 0.45 NA optical objective was employed.

As is seen in Figure 6A the polymer substrate has peaks at $600cm^{-1}$ and $814cm^{-1}$ which are essentially the same in the far-field and TERS near-field spectra. Thus, they act as a good internal intensity standard in comparison to the reported resonance enhanced spectra where the presence of the probe tip at this wavelength of excitation simply further enhances bands that are already seen in the far-field resonance Raman spectrum.

In contrast, the results at 785nm excitation (Figure 6A), which is out of the reported absorption profile of bilayer $MoS_2$ [29], show that there is little far-field Raman scattering other than the bands of the SU8 polymer substrate in this bilayer crystal. However, when the TERS tip is present an intense spectrum relative to the polymer scattering is observed and this spectrum has bands at $382cm^{-1}$, $408cm^{-1}$, $464cm^{-1}$ and $642cm^{-1}$. These bands are very close in frequency to those reported in the enhanced 660nm spectra [28].

It is interesting to note that, in a recent paper on bulk Raman scattering with 785nm excitation [30], a stronger spectrum is observed for the bulk material at this excitation frequency. This is probably due to the large interaction volume of the laser beam with the bulk material. In this spectrum a band close in frequency to $225cm^{-1}$ to the observed TERS band in Figure 6A is seen only with this far-field excitation.

In Figure 6B laser excitation wavelengths at 532nm and 514nm were used. These wavelengths are not in either of the excitonic absorptions as was the case with the reported 660nm results. Rather, they are in the rising high energy absorption background reported in reference 29. The resulting spectra are strongly resonance enhanced and as a result, unlike the 785nm excitation, little polymer scattering is seen. Also it is known from the bulk spectra that the bands observed at $385cm^{-1}$ and $408cm^{-1}$ which are also seen with 785 and 660nm excitation are related to stacking and the frequencies of the peaks are indicative of the number of stacked layers. Interestingly, once again the frequencies of these vibrations at different excitation wavelengths are essentially unchanged but the relative intensities show variation.

The above results together with those reported at 660nm [28] provide us in the near- (TERS) and far-field with four wavelengths of excitation in different regions of the absorption profile. In addition, four different substrates, three of which are non-plasmonic, namely the SU8 polymer, silicon and glass have been employed with the previously reported data at 660nm excitation being performed on a gold plasmonic substrate [28].

A general conclusion from all these results is that frequency of the vibrational modes are generally unperturbed in the data obtained over all these excitation wavelengths and with all these different surfaces but the intensity of the vibrational modes vary greatly.

*TERS of a non-resonant molecular system on a non-plasmonic surface*

The observed signal resulting from a gold ball TERS tip in contact with a supported bilayer membrane (SBM) sitting on a non-plasmonic mica substrate is shown in Figure 7. TERS spectra were obtained before and after the addition of lipid vesicles.

Distinct features are clearly observed in the TERS spectrum in the presence of lipids. Spectra obtained by retracting the tip from the surface show the features to be reproducible and dependent on the presence of the tip. Interestingly the observed TERS spectrum does not resemble the spontaneous Raman spectrum of the lipid. However, the spectrum associated with the tip has similarities with data observed by Taylor et al. [31] from a gold nanoparticle placed on an absorbed bilayer membrane formed on a Au surface. The bilayer membrane as reported by Taylor et al consisted of a self-assembled monolayer of alkanethiol with a lipid leaflet on top. The nanoparticles were then deposited on top of this bilayer without any AFM control.

Since the substrate is a mica di-electric surface, the effect observed is independent of the formation of a gap-plasmon mode as could be assumed in the Taylor et al results [31]. Careful inspection of the observed peaks suggests frequency shifts relative to those observed in the spontaneous Raman spectrum. The spontaneous Raman spectrum of DPPC shows the expected bands at 1737, 1440, 1298, 1128, 1099, 1066, 955, 925, 887, and 718 cm$^{-1}$.[32] However, in the TERS spectrum in the presence of lipids, significantly different bands are observed. The lipid bilayer TERS data shows bands at 1655, 1588, 1562, 1519, 1452, 1426, 1395, 1359, 1328, 1299, 1212, 1057, 961, 922, 821, 715, and 698 cm-1. In general vibrational modes at higher and lower frequencies are observed relative to the spontaneous Raman data of the lipid.

It has been shown that strong coupling between infrared vibrational modes and cavity modes result in frequency shifts [33]. Thus, the frequencies observed in the TERS spectrum of the lipids presented in this paper and in the results of Taylor et al [31] can be interpreted as strong coupling between the tip or gold particle with the lipids and its substrate.

This strong coupling has also been recently seen in the visible region of the spectrum at room temperature. In a seminal elegant paper Chikkaraddy et al [34] trapped 1-10 molecules of methylene blue using a gold nanoparticle on a gold mirror geometry combined with guest host chemistry to reduce the cavity volume to 40 cubic nanometers. Single methylene blue molecules were trapped in the cavity with their dipole aligned in the Z direction, the same as the plasmonic field. Under such conditions Rabi splitting in the scattering spectra have been observed. The only way to achieve such ultrasmall volume in scanned probe microscopy is either with tunneling feedback or AFM feedback with no jump to contact (as exemplified by tuning fork feedback). The gold mirror geometry essential for these experiments also explains why in TERS pristine gold surfaces with noble metal probes in tunneling feedback under ultrahigh vacuum are so effective at producing TERS.

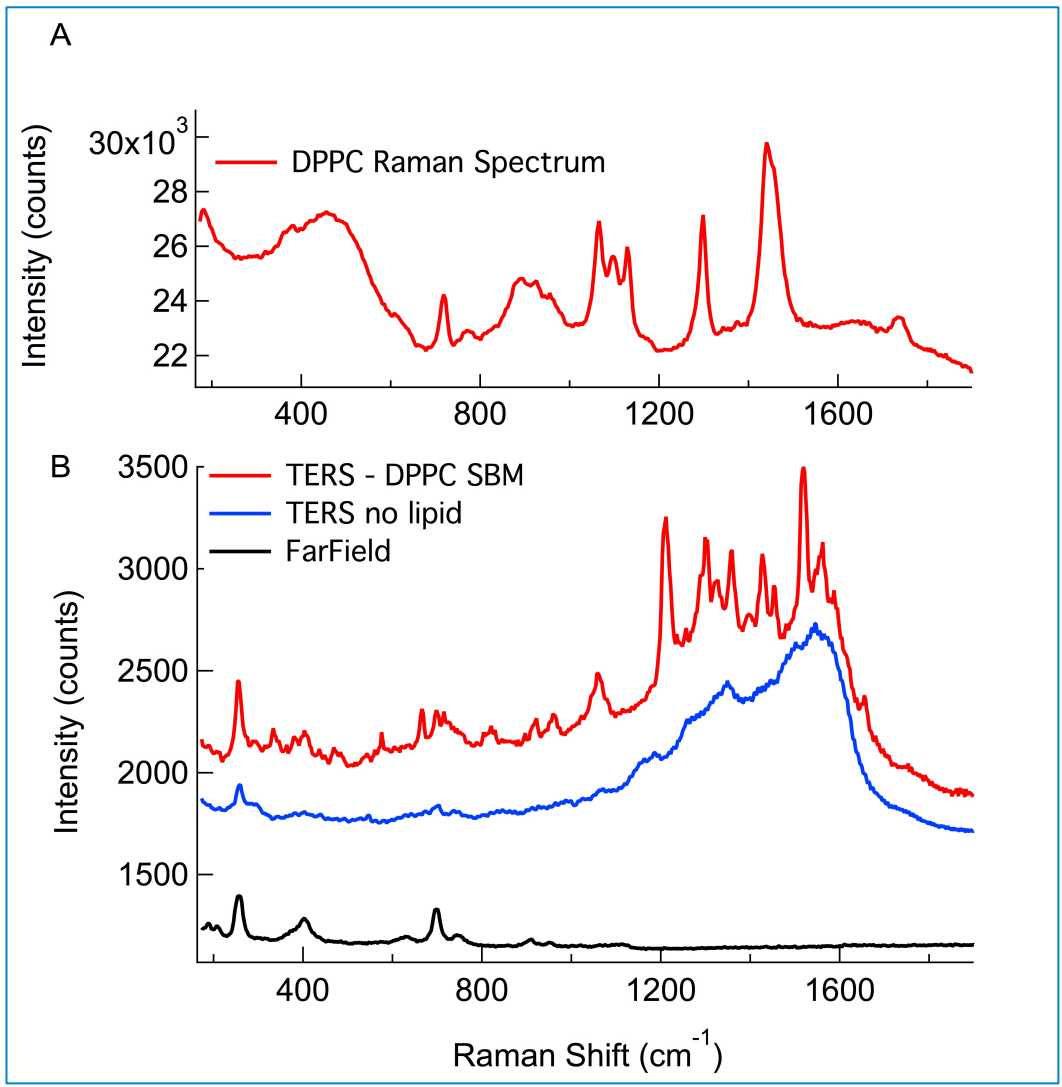

Figure 7. Panel A Shows the spontaneous Raman spectrum of DPPC vesicles in comparison to B that shows the TERS signals obtained from a DPPC supported bilayer membrane (red) and the spectrum in the absence of lipids (blue) formed on mica. The TERS spectrum of DPPC shows distinct differences from the spontaneous Raman spectrum of the lipid in A, which indicates that strong coupling between the tip and sample is occurring. The far-field background in B (black spectrum) is associated with mica.

Discussion

Considering the results reported in this paper and within the context of the body of literature that now exists on both TERS and on the quantum nature of near-field interactions it is possible to make a series of observations which suggest a new direction for understanding TERS.

The first observation is that the investigations reported in this paper on the possibilities of inducing plasmonic effects with a tunneling current with a gold tip on a gold surface are inconsistent with the important TERS results of Zhang et al [7]. As reported by these workers the voltages and currents associated with excellent TERS are one order of magnitude lower than what is needed for tunneling induced plasmons.

This is totally consistent and is supported by the recent results published by Chiang et al. [12]. In this study STM Induced Luminescence was compared with Tip Enhanced Fluorescence spectra. These workers show that, as in our study, to excite an electronic process, which in the case of Chiang et al is tunneling induced luminescence without laser excitation, both a higher bias voltage and tunneling current of 2.5V and 1000pA respectively are required. This is contrast to laser induced Tip Enhanced Fluorescence with tunneling feedback reported in the same paper with tunneling feedback conditions of 1.0 V and 300 pA.

Nonetheless, the nature of the tunneling feedback does indeed effect the TERS intensity (see Figure 5). This deduction is further supported by recently published data on TERS of carbon nanotubes with STM feedback [35]. Namely, the TERS intensity increases as the tunneling current increases such that the tunneling gap decreases. This is deduced from the tunneling equation (see supplementary material). Thus, this leads to a second observation which is the criticality of reducing the volume of the gap between the tip and surface.

However, as the tip/molecule/surface volume gets smaller one would expect that the field at constant voltage would increase. Such an increasing field should have a growing perturbing Influence on the vibrational spectrum of the molecular system trapped in the tunneling gap. Nonetheless, as a third observation it is noted that the large body of TERS and SERS results, including those reported in this paper, show that in most cases essentially no perturbation in the ground state vibrational frequency occur upon TERS or SERS enhancement.

This observation is also supported by the data of Zhang et al [7] and Liao et al [35] with a controlled quantifiable closer and closer approach of the tip to the surface that is achievable with tunneling feedback. Such controlled approach data show that the vibrational frequencies in the enhanced spectra are rigidly constant as the tip gets closer to the surface and the intensity is enhanced. This is further supported by the results of Chiang et al [12] showing no reproducible STM bias-dependence on the TERS signal.

This is also recently supported by the results on hexil azobenzene thiol monolayers on gold (111) which show essentially no frequency change as a function of voltage [36]. Interestingly in this case however, as the bias was increased a change in the orientation of the azobenzene tethered moiety with respect to the surface normal in the molecules located below the tip apex occured. The strong electric field eventually caused a tilting of the azobenzene plane(s) resulting in a decreased optical coupling and weaker TERS signal deviating somewhat from the distance dependence noted above. This is an important and a rare example of a well-documented effect of tunneling voltage on molecular geometry in TERS where the distance between the tip and surface is above 1nm as is the case in the examples described in references 7, 12 and 35.

On the other hand Chikkaraddy et al [34] achieve a distance of the tip to the surface of 0.9nm using guest host chemistry. This is critical for them to reach a strong coupling regime for a completely mixed matter light quantum state. This means that we can assume that TERS is occurring generally in a weak coupling regime >1nm where such mixed matter light quantum states are not achieved.

However, as already shown by Savage et al [37] at such distances above 1nm and <50nm near-field coupling occurs.  This coupling increasingly perturbs the optical absorption of the tip/sample complex as smaller distances are achieved between the tip and the substrate.  This perturbation of the optical absorption on the close approach of two surfaces has been experimentally demonstrated by the Savage et al [37].

Nonetheless, in spite of the clear increase seen in the TERS intensity in the near-field weak coupling regime, Savage et al [37] note that when the tunneling regime of a few nm is reached the phenomenon of tunneling opens a conductance channel between the tip and the surface, modifying the charge distribution and reducing the strength of the plasmonic interaction.  But this is opposite to what would be expected to be observed for a purely plasmonic interaction affecting the TERS intensity which increases as the tunneling distance decreases.

This reduction in the plasmonic interaction at the onset of tunneling is also supported by the theoretical results of Maranica et al [38].  These workers show that the capacitive coupling between the nanoparticles at these distances are attenuated as tunneling is established and the corresponding plasmonic modes progressively disappear upon the narrowing of the gap.

Thus, in addition to plasmonic interactions, some other spectral phenomena must underlie the basis of the TERS intensity enhancements as the volume sandwiching the molecule between a tip and a surface is progressively reduced.

In fact it is clear from the literature that as a tip and a surface approach one another new hybrid state absorptions are revealed both from a theoretical [39, 40] and experimental perspective [34, 37].  Thus, as a function of the cavity gap, the optical response of the tip/sample/substrate complex is altered in the near-field regime. Furthermore, for such alterations in optical response, the presence of noble metals are not essential.  In fact, such hybrid quantum state formation with modified absorptions can occur even with silicon dimers as shown in supercontinuum absorption studies with silicon dimer nanoparticles [41].

Searching for previous examples of similar vibrational mode intensity enhancements with essentially no ground state vibrational frequency alterations one is led to resonance Raman spectroscopy.  Resonance Raman spectroscopy shows enhancements similar to TERS.  The extent of enhancement depends on the position of the Raman excitation wavelength relative to a vibronic transition with many examples of pre-resonance effects for excitations that are even far from exact resonance.

In the case of TERS the hybrid state absorptions due to the separation of the tip and surface distance sandwiching the molecule could vary the coupling of the excitation laser with the sample.  The lack of frequency alterations in contrast to significant intensity variations could readily be explained in a resonance Raman based TERS intensity enhancement mechanism associated with such hybrid state absorptions. This is clearly seen in the case of the $MoS_2$ resonance Raman spectra reported in this paper where the nature of the TERS results depend on the wavelength of the excitation laser vis a vis what could be complex absorptions as the tip sandwiches the molecule to the substrate. For example, the data indicate that the presence of the tip induces new light sample interactions at 785nm (see Figure 6A) which is out of resonance in the far-field giving rise to similar molecular vibrational modes reported by Zhang et al [28] at 660nm excitation which is in resonance.

The defining factor in all such enhancements in a resonance Raman context would be the Frank-Condon overlap between the ground and the excited state for the particular excitation wavelength chosen relative to the hybrid state absorption. The degree of overlap is predicated by the offset of the excited virtual state relative to the ground state. Within this context, in TERS the frequency of the Raman excitation laser relative to the hybrid absorption of the tip/ molecule/ substrate sandwich complex has in the past not been considered. Thus, fixing the laser excitation wavelength independent of this understanding could readily account for considerable variability in the effectiveness of TERS.

In fact, resonance Raman scattering has been invoked previously by Van Duyne et al. to explain intensity fluctuations observed at a single excitation wavelength [42] resulting in single molecule TERS fluctuations. And Lombardi et al [43] have observed the alterations in the absorption profile by the presence of the tip due to quantum interactions with the surface.

The theoretical basis of understanding such resonance Raman scattering was pioneered by Albrecht [44] with a recent emendation reported by Gong et al. [45]. A qualitative illustration is shown in Figure 8. In accordance with this illustration, when a laser interacts with a quantum state there is a vertical transition to a virtual state that is defined by the Frank-Condon overlap. Second order perturbation theory forms the theoretical underpinnings of this understanding of resonance Raman scattering. Within this understanding the intensity of the Raman spectrum depends on the polarizability tensor which is related to the Raman tensor. The Raman tensor is composed of a series of matrix elements and a resonant denominator. The matrix elements describe the coupling of the vertically excited virtual state (shown as excited by the blue arrow in Figure 8) to the state from which coupling back to the ground state occurs (green arrow). The principle component in this transition is the change in the Hamiltonian with respect to the normal coordinate that is correlated with this excited state coupling. The specific normal coordinate involved is the vibrational frequency in question that has its intensity significantly altered but its frequency remains the same as that in the ground state. This intensity enhancement is strongly related to the excitation wavelength and the associated Frank-Condon overlap that defines the vibronic states involved.

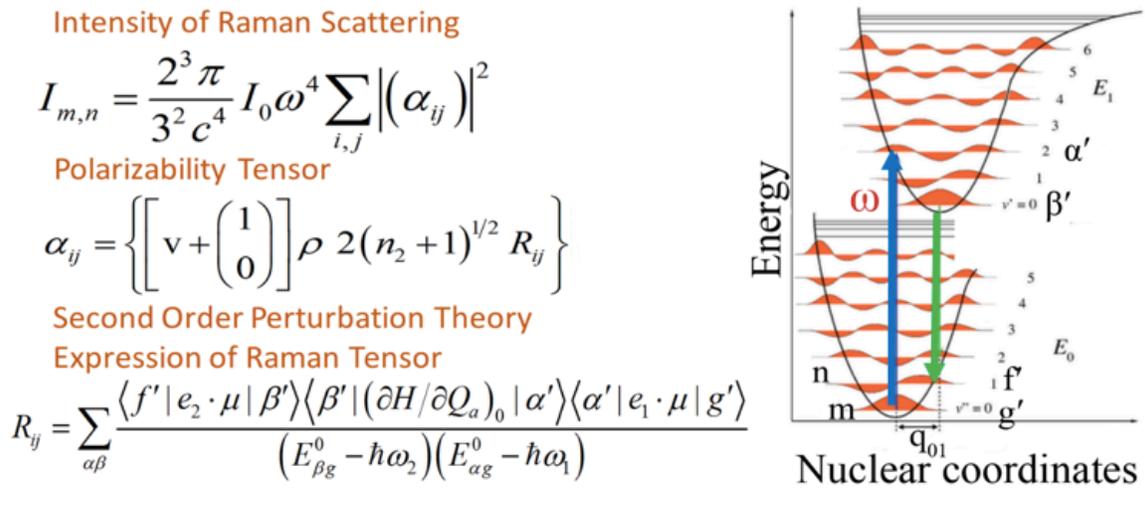

Figure 8. Qualitative illustration describing, in concert with the equations shown, the theory of resonance Raman Scattering.

Within such a framework, tip induced perturbations in the absorption profile [42,43] would be considered as a weak coupling perturbation which would be associated with molecules that are in or near-resonance with the Raman excitation wavelength. As noted above, an example of such weak coupling are the TERS results on the $MoS_2$ bilayer crystal reported in this paper. In contrast, to detect TERS of the bilayer lipid system, which is very far from resonance, one needs a strong perturbation of the molecule being sandwiched between the tip and the substrate. As a result pronounced alterations in the ground state vibrational frequency are detected and this has been called a strong coupling interaction [31-34] regime .

From a perspective of such weak and strong coupling the alterations in TERS relative to far-field spectra the framework introduced by Chikkaraddy et al [34] is very helpful with regard to understanding deeper the TERS effect. As noted by these workers to enhance coupling strengths the effective cavity volume has to be reduced. Furthermore, the coupling strength is directly related to the transition dipole and its orientation.

Chikkaraddy et al [34] have shown that the smallest cavity volumes and controlled dipole orientation can be obtained with nanoparticle mirror geometry. With this approach a gold nanoparticle on a gold surface sandwiches a molecule with controlled dipole orientation with the aid of guest host chemistry.

Based on these data and the associated understandings provided by Chikkaraddy et al [34] the relative success of STM TERS with noble metals or metal nanoparticles approaching a surface can be better appreciated. Coupled to this is the criticality of the dipole of the transition that is to be enhanced by the presence of the tip. The inter-relation between the dipole and the cavity volume could also in part explain why coupling laser excitation for TERS can result in variability in results often seen.

A good example of such variability is the appearance of the Amide I mode in TERS and SERS spectra of proteins [46,47]. This mode is associated with the peptide bond that has an absorption at 193nm far from visible laser used in TERS. Thus, both the geometry of this structure relative to the tip/sample/sandwich and its dipole orientation would be crucial to both the hybrid absorption and the coupling of the visible laser employed for TERS excitation. Related to this are the results Kuroski et al [47]. Their results indicate that bulky amino acid side chains could be preventing effective coupling to the peptide bond. In contrast un-shifted aromatic amino acid modes are prominent [46] in such protein TERS spectra and this could be due to and the strong pre-resonant behavior of these moieties even when visible laser excitation is employed.

All of the above observations, summarized below, suggest that resonance Raman phenomena could well be an underlying fundamental mechanism controlling TERS enhancement.

Observations Supporting A Fundamental Role For Resonance Raman Enhancement In TERS

- ➤ TERS intensity increases with reduced tip substrate volume when either tunneling or other feedback mechanisms [7,12,28] are employed while the plasmonic field is reduced when tunneling is established [37,38] inducing plasmonic modes to disappear [38].

- ➤ TERS and SERS results are independent of the use of noble metals for tip and substrate materials [19] and examples of SERS substrate viability has been demonstrated even when no noble metals are involved [20,21].

- ➤ Perturbed optical absorptions and hybrid states occur in both noble metal and semiconductor materials that are closely approached [41].

- ➤ TERS spectral effects, like those in resonance Raman enhancements, show essentially no alterations in ground state vibrational frequencies.

- ➤ In spite of the lack of frequency shifts large intensity alterations are seen in TERS and such variations are consistent with excited state effects due to alterations in Frank Condon overlap which have been seen to dominate resonance Raman enhancements

To fully test this hypothesis one needs to combine a supercontinuum source on-line with a TERS system. This will allow the effective correlation of TERS enhancements with the detected absorption alterations induced by a probe as it approaches within close proximity with a surface. The first steps in such measurements have just been published by Sanders et al [48] who applied a supercontinuum source to investigate the absorption characteristics of three different production methods for noble metal tips. Such measurements have to be extended to the complete TERS geometry to better understand and interpret TERS experimental successes and failures.

Summary

The present work reports several different streams of evidence suggesting that an underlying general mechanism of what has been seen in TERS enhancements of vibrational mode intensities occurs due to weak coupling induced, tip generated vibronic alterations affecting Frank-Condon coupling into virtual excited states.  This leads to the nearly general phenomenon of intensity enhancements with essentially no ground state vibrational frequency alterations. Resonance Raman scattering is the most likely explanation for this form of frequency independent intensity enhancement in resonant and near resonant systems. Our results emphasize the close approach of a tip to a surface as being the fundamental perturbation, which, through quantum interactions of the tip with the surface, alters the coupling of light with a molecule to give rise to the resonant enhancement of vibrational modes without frequency alterations. For non-resonant systems strong coupling is invoked that leads to enhancement of the intensity with frequency alterations.

**Acknowledgment.**

ZDS acknowledges support from the National Science Foundation award CHE-1507287 and the National Institute of General Medical Sciences (NIH) Awards R00 RR024367 and R01 GM109988. JP acknowledges financial support from the GSRT of the Ministry of Education, Research and Religious Affairs, Greece in the frame of Greece-Israel collaboration for Joint R&D under the "Regions at the center of development" co-financed by the Hellenic Republic and the EU (Project 3100-2DNanomechanics). AL acknowledges support from the Israel Ministry of Defense.

**Supplemental Material**

The graph in Figure 5 which plots the TERS intensity as a function of the tip substrate distance is based on the results of Zhang et al [7].  The spectra obtained by these workers as a function of tunneling distance of a tip from the surface at constant voltage provided the data that was extracted from Figure 2(c) of the results of Zhang et al [7].  For the extraction the Matlab Program "Grabit" [https://www.mathworks.com/matlabcentral/fileexchange/7173-grabit] was employed.  The most intense band in the spectrum of meso-tetrakis(3,5- di-tertiarybutylphenyl)-porphyrin at ~1210 cm-1 was chosen for the graph in Figure 5 although other vibrational modes would have given similar results.  The values on which the graph was based are the following: At 1pA of current the band was estimated to have a Raman shift at $1214cm^{-1}$ and a relative intensity of 1888.  At 10pA of current the band was estimated to have a Raman shift at $1207cm^{-1}$ and a relative intensity of 2436.  At 50pA of current the band was estimated to have a Raman shift at $1205cm^{-1}$ and a relative intensity of 3613.  At 100pA of current the band was estimated to have a Raman shift at $1206cm^{-1}$ and a relative intensity of 4288.  At 300pA of current the band was estimated to have a Raman shift at $1209cm^{-1}$ and a relative intensity of 5921.  At 500pA of current the band was estimated to have a Raman shift at $1213cm^{-1}$ and a relative intensity of 8299.

Based on this data and given that the tunneling current is known to be proportional to the distance between the tip and the surface with a functional dependence of:

$$I(d) \propto V_{bias} \exp(-kd)$$

where $k = \sqrt{\dfrac{8m\Phi}{h^2}}$, m is the electron mass and Φ the work function of silver (the metal employed by Zhang et al [7]) a graph could be constructed of the TERS intensity as a function of distance of the tip to the substrate. For such a graph displayed in Figure 5 the applied bias voltage reported by Zhang et al [7] of 120mV was employed to determine the distance between the tip-surface using $d \propto 1/k \ln(V_{bias}/I)$.